\documentclass[prb,preprint,superscriptaddress,aps]{revtex4-2}
\usepackage{amsmath}  
\usepackage{amsfonts} 
\usepackage{graphicx} 
\usepackage{url}
\usepackage{xcolor}
\usepackage{natbib}
\graphicspath{{Images/}}

\begin{document}

\title{Integrating Evolutionary Biology into Physics Classroom: Scaling, Dimension, Form and Function}

\author{Kausik S Das}
\email{kdas@umes.edu}
\affiliation{Department of Natural Sciences, University of Maryland Eastern Shore, 1, Backbone Road, Princess Anne, MD 21853 USA}
\author{Larry Gonick}
\affiliation{247 Missouri Street, San Francisco, CA, 94107}
\author{Salem Al Mosleh}
\affiliation{Department of Natural Sciences, University of Maryland Eastern Shore, 1, Backbone Road, Princess Anne, MD 21853 USA}

\begin{abstract}
Since Galileo and (more recently) D’Arcy Thompson, it has been understood that physical processes and constraints influence biological structures and their resulting functions. However these cross-discpline connections --- and their importance to growing scientific discplines such as biophysics ---  are rarely tought in introductory physics courses. Here we examine how the laws of physics shape Darwinism evolution through the surface area to volume ratio, an important geometric measure of a structure. We develop conceptual cartoon clicker questions to enhance students' understanding of these interdisciplinary concepts. By connecting abstract physical laws with biological (and technological) applications, our approach aims to help students appreciate the deep connections between physical and biological sciences, thereby enriching the learning experience in introductory physics courses.
\end{abstract}

\maketitle 

\section{Introduction}
Physics is the study of quantitative laws that describe the natural world. While these laws apply equally to living and non-living systems, early physics education focuses mostly on the latter, neglecting rich connections with biology. Galileo's observation~\cite{galilei1914dialogues} --- suggesting that extremely large organisms cannot have the same proportions as smaller ones without collapsing under their own weight --- highlights an early example of the relation between physics and evolutionary biology. Three centuries later (1917), D'Arcy Thompson~\cite{d1917growth} described the rich interplay between biology, physics, and mathematics in determining how an organism's shape emerges during both evolution and development. Building on Thompson's pioneering ideas, Recently scientists extended and applied this approach to diverse living systems~\cite{chelakkot2017growth, tallinen2016growth, savin2011growth, blackie2024sex, irvine2017mechanical, flatleaf, al2021geometry, bookstein2013measurement, heisenberg2017d}.

Neo-Darwinism posits that genes mutate randomly, with natural selection favoring advantageous mutations \cite{jarron2023single, kutschera2004modern}. Mutated genes result in new species, and those mutations that provide survival advantages in a given environment become dominant. The environment, in turn, is governed by the physical laws of nature. Thus, adaptations are influenced by these natural laws of physics.

The surface area to volume ratio is an important quantity in fields such as biology, chemistry, and engineering, where surface effects can significantly influence the behavior, function and properties of materials and organisms at small length scales~\cite{harris2018surface,das2020microbial,pan2008semiconductor}. The article ``On Being the Right Size" by J. B. S. Haldane~\cite{haldane1926being} explains that as animals increase in size, their surface area grows at a slower rate than their volume. This affects various aspects of their physiology and capability, such as bone strength, muscle power, and the efficiency of their respiratory and circulatory systems. He uses different examples from the animal world highlighting the physical limitations imposed by the surface area-to-volume ratio. Erwin Schrodinger --- in the beginning of his classic book ``what is life'' \cite{schrodinger1992life}, which discusses the role of physics in biology --- discusses the role that size of organisms and their sensing apparatuses plays in allowing for robust predictable behavior.
In this paper, we illustrate how physical laws determine evolutionary adaptations of various species through the surface area to volume ratio, a quantity that depends both on size and shape of an organism. This will also demonstrate the principle that form follows function~\cite{russell1916form}, which is manifested across scales in the living world, from protein shapes to ecological communities like coral reefs. Consistent with the National Science Education Standards \cite{national1996national,taylor2009proportional}, this paper discusses integrating growth, form, and evolutionary biology into early physics education. We propose using concept cartoon clicker questions to enhance student understanding by encouraging student engagement. 
This approach embodies the recommendations of the National Research Council that teaching introductory physics should emphasize active student participation\cite{national2013adapting}.

This paper is organized as follows. In Sec.~\ref{sec:define-ratio}, we introduce the surface area to volume ratio using various geometric objects as examples. This section and associated cartoon clicker questions provide intuition for how this ratio changes with size and shape of a system, and how it increases when an object is fragmented. In section ~\ref{sec:evolutionary-bio}, we discuss how the surface area to volume ratio affects evolutionary adaptations of species. Examples, include the effect of the ratio on nutrient and oxygen diffusion, heat loss, and mechanical load bearing capacity. Then, in section ~\ref{sec:technology}, we discuss the role of the surface area to volume ratio in technological applications such supercapacitors and heat exchangers.

\section{Understanding surface area to volume ratio} \label{sec:define-ratio}

The surface area to volume ratio, $\rho \equiv A/ V$, is a geometric property, with dimensions of inverse lengths, that characterizes the shape of an object and scales inversely with characteristic size. For example, the size of a cube is characterized by its length $l$ and the surface area to volume ratio is calculated as follows,
\begin{eqnarray}
    A_{cube} = 6 l^2, \;\;\;\;\;\; V_{cube} = l^3 \;\;\;\;\implies \rho_{cube} = \frac{6 l^2}{l^3} = \frac{6}{l}.
\end{eqnarray}
We see that $\rho_{cube}$ decreases linearly with its size and the factor of 6 is result of the particular shape of the cube. For a different objects, such as sphere whose size is characterized by the radius $r$, we have
\begin{eqnarray}
    A_{sphere} = 4\pi r^2 \;\;\;\;\;\; V_{sphere} = \frac{4}{3}\pi r^3 \;\;\;\;\implies \rho_{sphere} = \frac{4\pi r^2}{\frac{4}{3}\pi r^3} = \frac{3}{r}.
\end{eqnarray}
Here $\rho_{sphere}$ is inversely proportional to the size of the sphere, as in the case of a cube but with a different coefficient (3 in this case).

The choice of the characteristic scale is not unique. For example, considering a cylinder of radius $r$ and height $h$, we can choose either one as characterizing the size of the system. Here will take $r$ as defining the size of the system, while the aspect ratio $n \equiv h/r$ will characterize the shape of the cylinder. To calculate the surface area of the cylinder, we assume the cylinder is capped by two disks on both of its ends, which gives
\begin{eqnarray}
    A_{cylinder} &=& 2\pi rh + 2\pi r^2 = 2\pi r(nr) + 2\pi r^2 = 2\pi r^2(n+1), \nonumber \\
    V_{cylinder} &=& \pi r^2h = \pi r^2(nr) = n\pi r^3 \implies \rho_{cylinder} = \frac{2\pi r^2(n+1)}{n\pi r^3} = \frac{2(n+1)}{n}\cdot\frac{1}{r}.
\end{eqnarray}
This result shows explicitly how $\rho_{cylinder}$ depends both on the size of the system through $r$ and its shape through $n$. In the following, we will denote the generic variable for the size of the system as $L$, which can represent the the length of a cube $l$, or radius $r$ for cylinders and spheres.

\begin{figure}[ht!]
\centering
\includegraphics[width=6.0in]{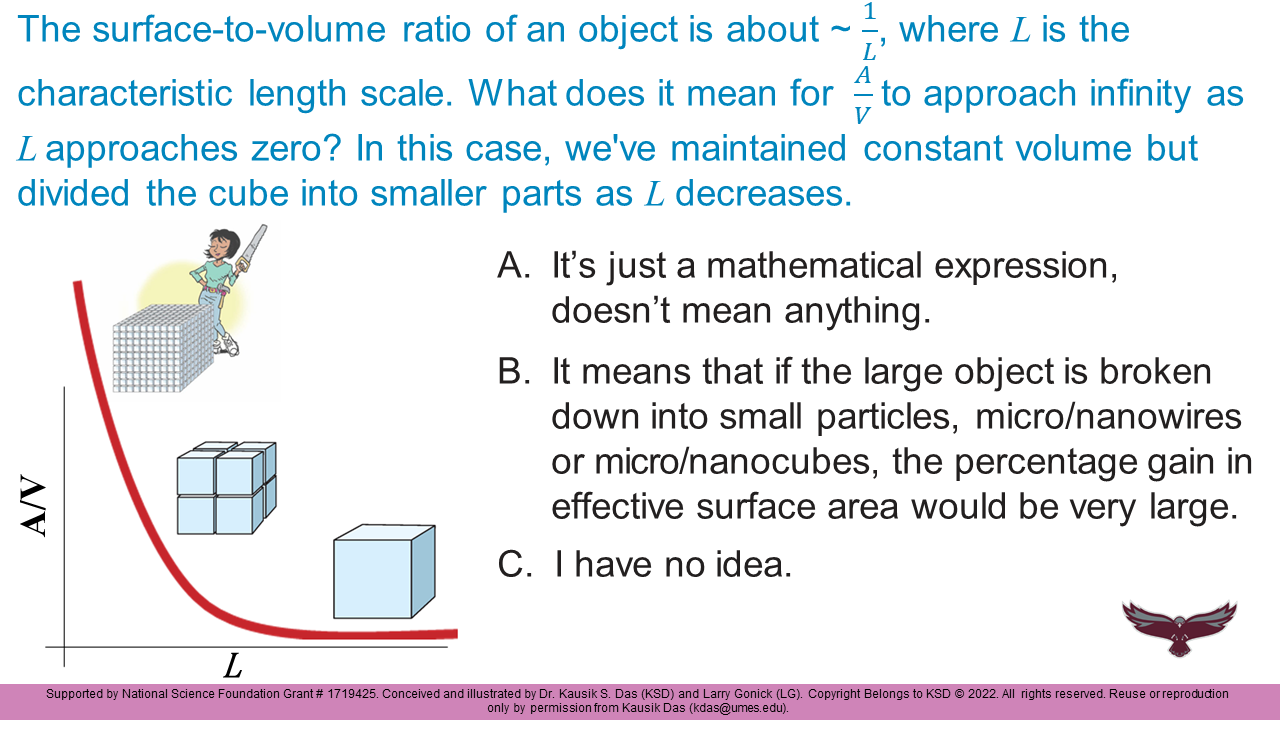}
\caption{\textbf{Surface area to volume ratio (A/V) decreases with typical size of an object}. This question illustrates this concept by breaking down a large cube (which has a relatively small A/V) into smaller pieces with smaller characteristic length scale, thus generating excess surface area without affecting the total volume. }
\label{A_V_Asymptotic}
\end{figure}

\begin{figure}[ht!]
\centering
\includegraphics[width=6.0in]{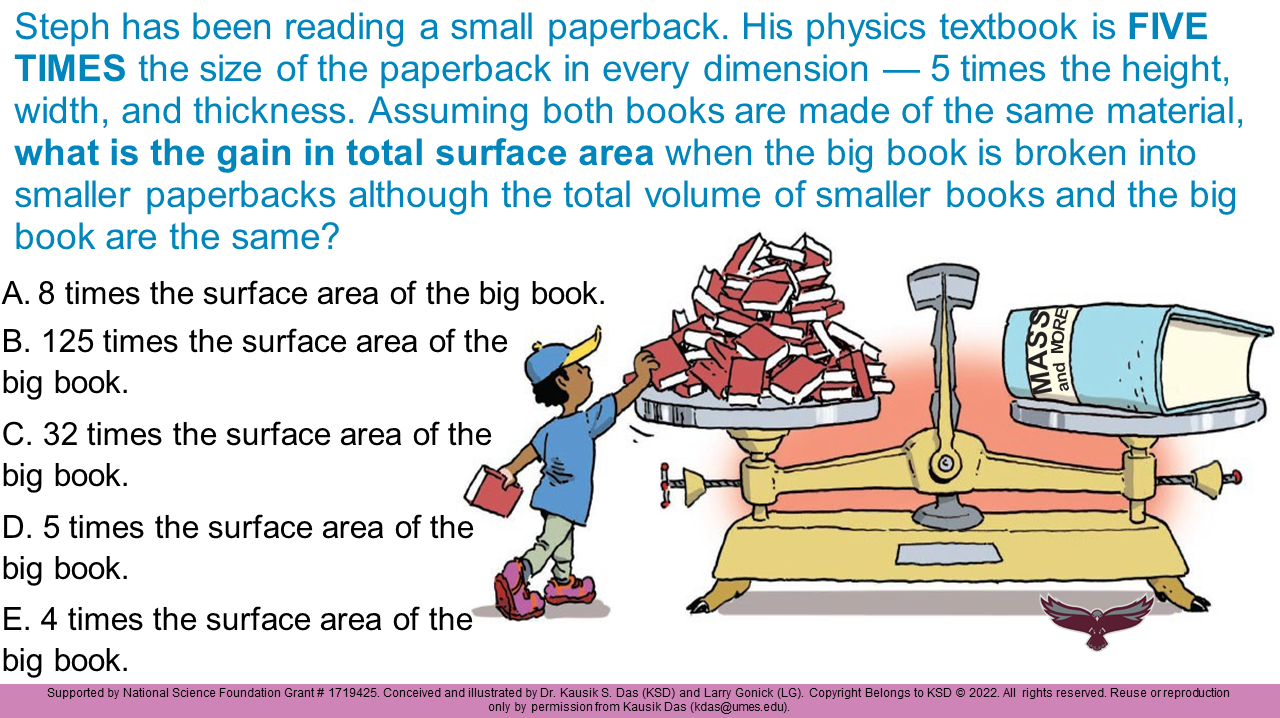}
\caption{\textbf{Surface area to volume ratio and fragmentation}. This question further illustrates the concept in Fig.~\ref{A_V_Asymptotic} by considering a large object (book), which is compared to several small books whose combined volume equals that of the book. Assuming all the small books have the same aspect ratio, the total surface area will be $5^3 = 125$ times for the smaller books compared with that of the single large book whose side lengths are 5 times larger than the smaller books.}
\label{Book_SA}
\end{figure}

As the object's size expands (i.e., the characteristic length $L$ increases while preserving the shape), the A/V ratio diminishes, and conversely, it rises when the size diminishes, as depicted in Fig.~\ref{A_V_Asymptotic}. The two extreme conditions, when the characteristic length $L$ approaches 0 or $\infty$, illustrate that surface effects become more pronounced at smaller sizes, while bulk properties tend to overshadow surface phenomena at larger sizes. This concept is further elaborated upon through the subsequent analysis that shows splitting a book into multiple smaller volumes increases the combined surface area while the total volume (and mass) remains the same (see Fig.~\ref{Book_SA}).

\begin{figure}[ht!]
\centering
\includegraphics[width=\columnwidth]{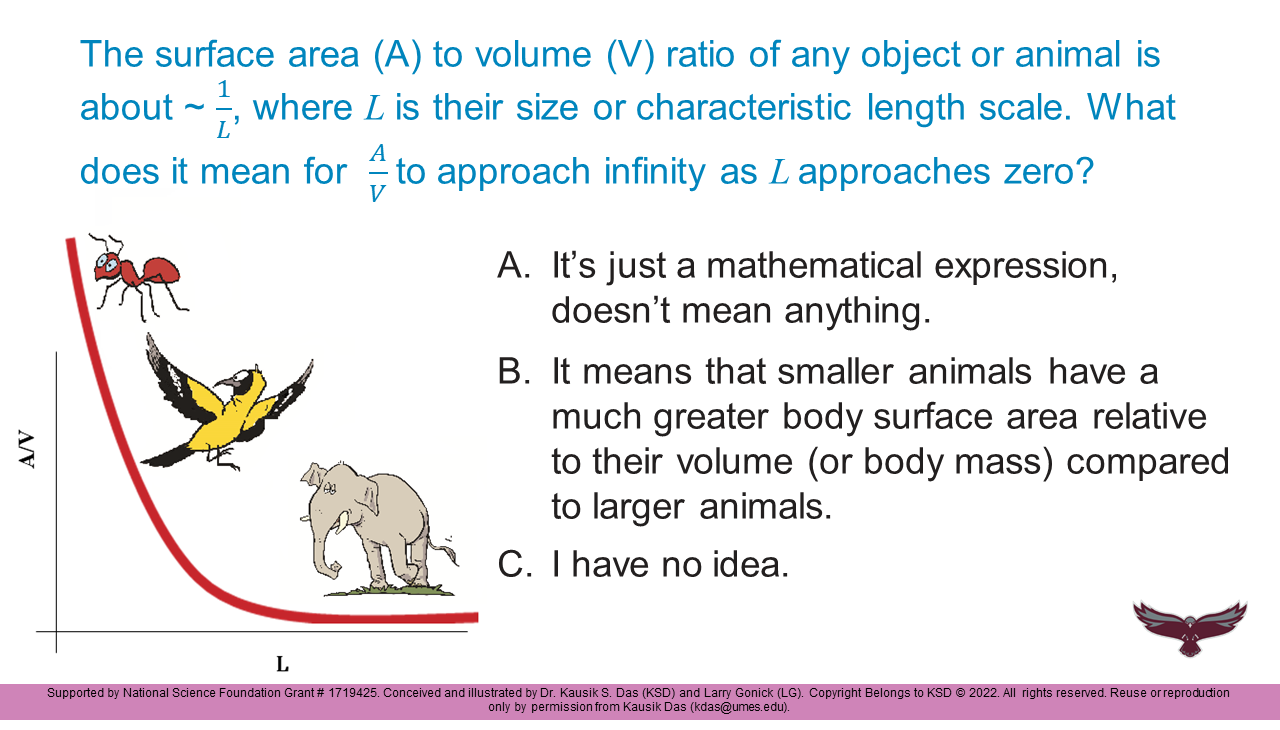}
\caption{\textbf{Surface area to volume ratio in natural world}. This question illustrates how smaller characteristic length scales impact the total surface area per unit volume (or body mass) in nature. It shows that an ant's surface area relative to its weight is greater than that of a bird, and significantly greater compared to an elephant.}
\label{Animal_size}
\end{figure}

Figure \ref{Animal_size} demonstrates that in the natural world, as the size of an animal decreases, the surface area-to-volume ratio of its body increases significantly compared to that of a larger animal. The examples below will illustrate this concept and its implications for biological evolution.


\section{Evolutionary Biology} \label{sec:evolutionary-bio}

\subsection{Oxygen Diffusion and Respiratory System}
The availability of sufficient oxygen to cells is a crucial factor for the survival and growth of organisms, playing a key role in the evolution from single-celled to multicellular life forms~\cite{bozdag2021oxygen}.


\begin{figure}[ht!]
\centering
\includegraphics[width=6.0in]{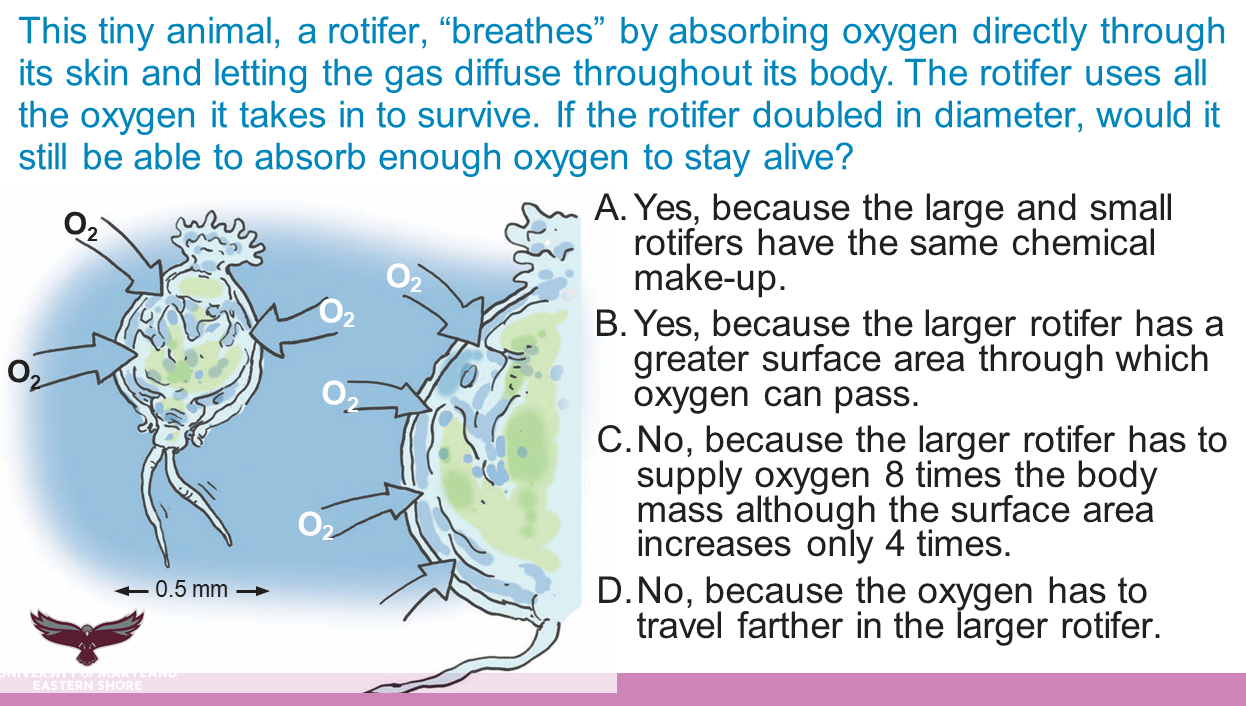}
\caption{\textbf{Diffusion limitation, breathing, and surface area to volume ratio.} The illustration shows how a rotifer breathes by absorbing oxygen directly through its skin, allowing the gas to diffuse throughout its body. The rotifer relies on this oxygen intake for survival. The question posed is whether a rotifer would still be able to absorb enough oxygen if it doubled in diameter. The correct answer (C) highlights the challenge that as the rotifer's size increases, its body mass increases at a faster rate than its surface area, making oxygen diffusion less efficient.}
\label{Rotifer}
\end{figure}

Fick's law of diffusion through a membrane describes how the rate of diffusion is proportional to the surface area (A), the concentration gradient (\(\Delta C\)), and inversely proportional to the thickness of the membrane (d):
\begin{equation}\label{DiffEq}
\text{Rate of diffusion} = D \frac{ A \cdot \Delta C}{d},
\end{equation}
D being the proportionality constant known as the diffusion coefficient.

According to this law, an increase in surface area (A) will result in an increased rate of diffusion, which is critical for small organisms that rely on oxygen diffusion through skin for respiration and other vital processes. For example, rotifers, which are microscopic aquatic animals as shown in Fig.~\ref{Rotifer}, have a unique respiratory mechanism where they absorb oxygen directly through their skin, allowing the oxygen to permeate throughout their body and supply sufficient oxygen to each of its cells to sustain vital functions. This method of respiration is efficient for their small size, as the surface area of their skin relative to their body volume is sufficiently large to absorb the oxygen they require. Considering the scenario where a rotifer's diameter is increased by orders of magnitude via genetic mutations, the implications for its survival become a matter of geometric scaling. The volume of the rotifer, which determines the amount of oxygen needed, scales with the cube of the radius (or diameter). However, the surface area through which the rotifer absorbs oxygen only increases with the square of the radius (or diameter). This disparity between the rate of increase in surface area and volume means that the larger rotifer would not have sufficient surface area to absorb the amount of oxygen needed for its greatly increased volume. The oxygen requirements of the body cells of the rotifer would outpace its oxygen absorption capabilities, leading to an insufficient supply of oxygen to maintain its metabolic functions. The larger rotifers would struggle to survive and eventually become extinct.

\begin{figure}[ht!]
\centering
\includegraphics[width=6.0in]{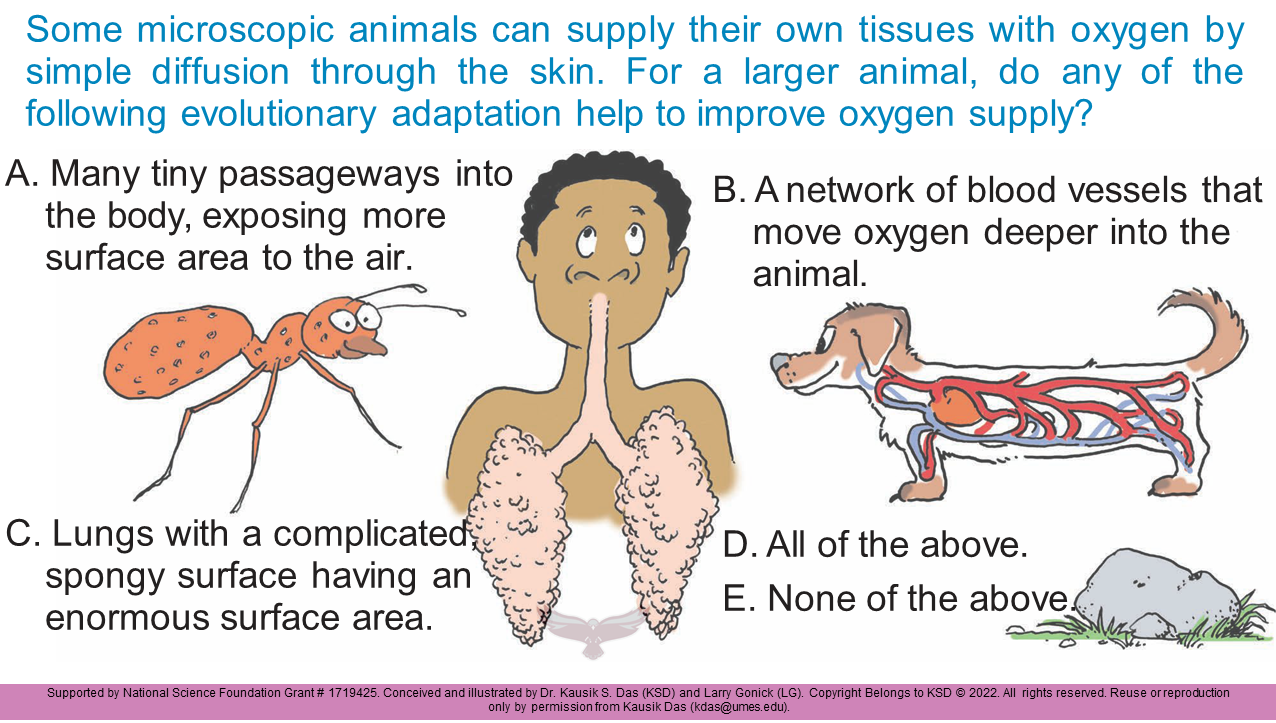}
\caption{\textbf{The need for lungs and circulatory systems motivated by the surface area to volume ratio.} This illustration explores the evolutionary adaptations that enable larger animals to improve oxygen supply compared to microscopic animals that rely on simple diffusion through the skin. The correct answer (D) highlights the various biological adaptations that have evolved to support efficient oxygen transport in larger organisms, compensating for the relatively smaller increase in skin surface area compared to their mass.}
\label{Oxygen_Supply}
\end{figure}

\begin{figure}[ht!]
\centering
\includegraphics[width=6.0in]{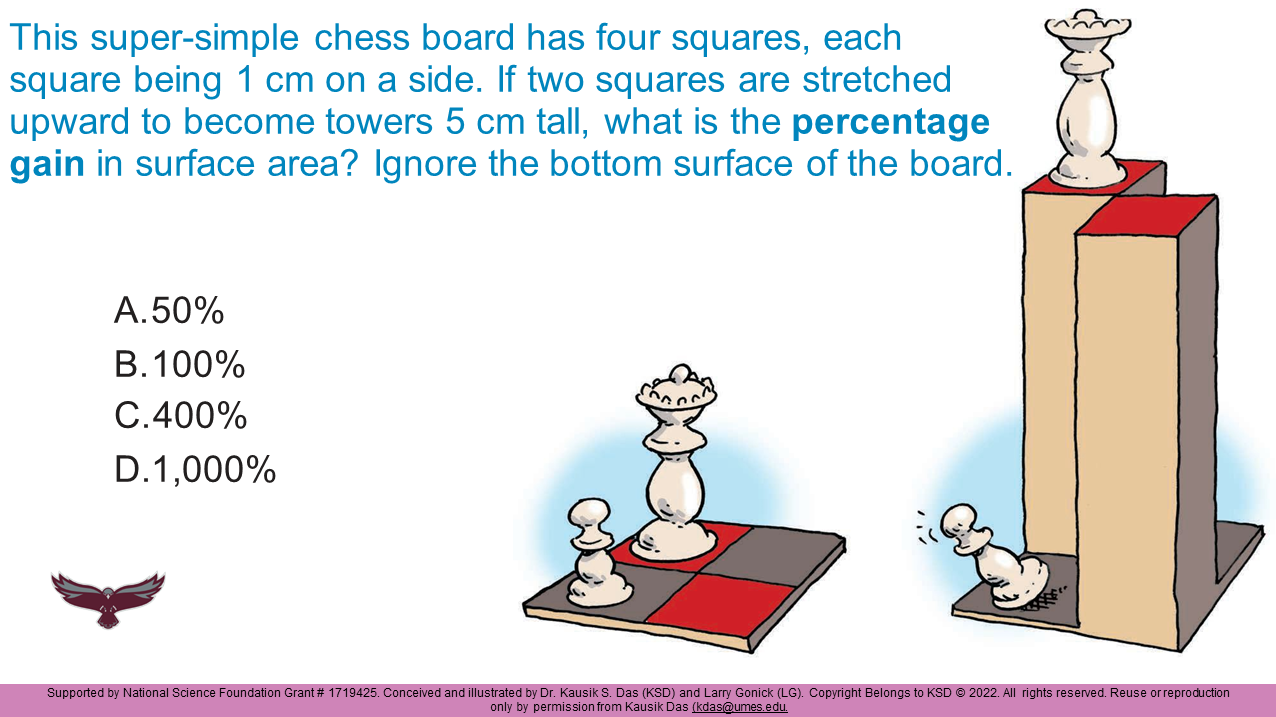}
\caption{\textbf{Illustrating the effect of surface topography on surface area.} A chessboard is transformed by stretching two squares into tall towers. This example calculates the percentage gain in surface area --- and resulting increase in surface area to volume ratio --- in topographic surfaces contrasted to smooth surfaces ($400\%$).}
\label{Chess}
\end{figure}

\begin{figure}[ht!]
\centering
\includegraphics[width=6.0in]{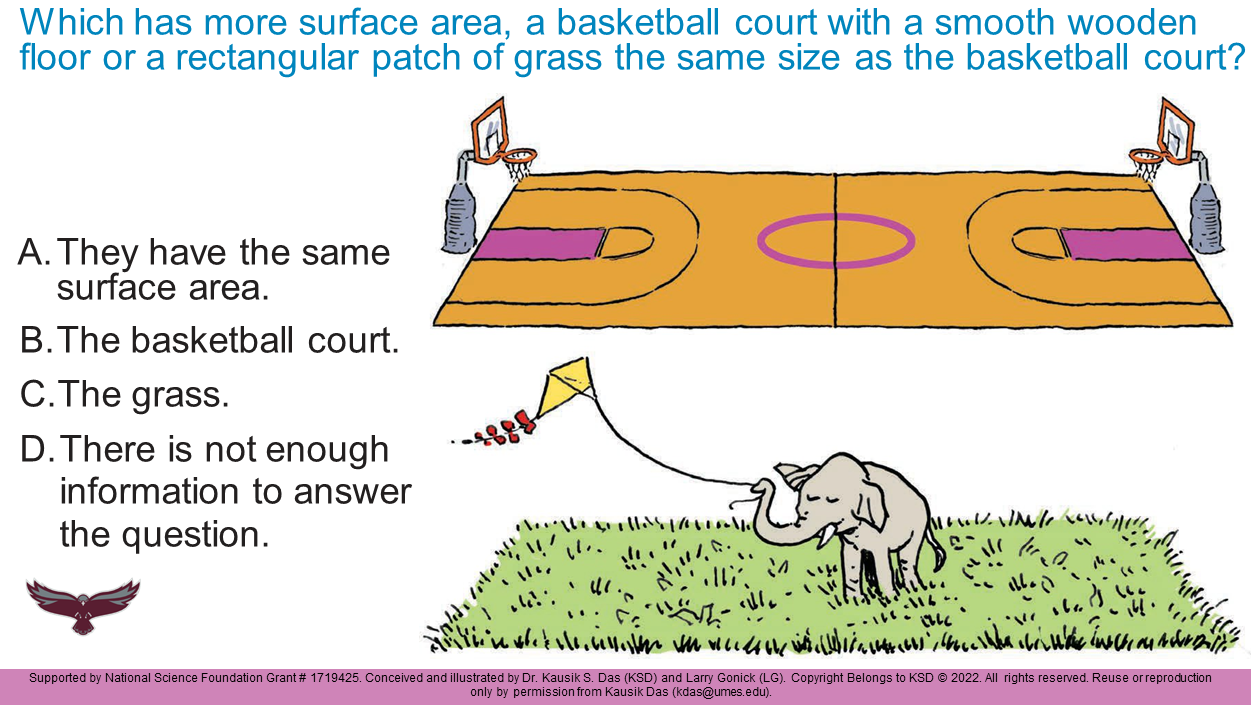}
\caption{\textbf{Effect of roughness on surface area.} A basketball court with a smooth wooden floor is contrasted with a rectangular patch of grass of the same horizontal size. The question explores which has a greater surface area, emphasizing the impact of texture and surface features on overall gain in surface area.}
\label{Field_Elephant}
\end{figure}

Therefore, during the course of evolution, as organisms grow in size, their respiratory systems must also evolve to supply oxygen to a vast array of cells, as shown in Fig.~\ref{Oxygen_Supply}. Natural selection favors genetic mutations that enhance oxygen uptake and distribution by creating additional surfaces. This creation of additional surfaces can occur through changes in surface topography, roughness, and other modifications, as illustrated by Figs. ~\ref{Chess}-\ref{Field_Elephant}. This mechanism is often seen in the development of lungs with a spongy texture, which greatly increases the surface area for gas exchange, and in the complexity of blood vessel networks that ensure oxygen is distributed throughout the body. In species like fish that inhabit water, gills are essential for breathing, allowing them to extract oxygen from their aquatic environment. Gills are made up of a large number of gill filaments, which have a large surface area to volume ratio, allowing for efficient gas exchange. These adaptations are vital for the survival of larger animals, as they enable efficient breathing and meet the heightened metabolic demands. Figs.~\ref{Chess}-\ref{Field_Elephant} provide clearer insight into how texture or topography contributes to the increase in surface area. These illustrations demonstrate that the introduction of roughness or varied topography (like gills) through mutations results in a more extensive surface area and better chance of survival.

It is clear that these evolutionary traits have been crucial for larger animals to overcome the limitations imposed by the surface area to volume ratio, where volume increases faster than surface area as an organism grows. With specialized respiratory structures as a result of random mutations that increase the surface area for oxygen absorption, these animals have been able to maintain efficient respiratory systems throughout their evolutionary changes, surviving by aligning themselves with the physical laws of nature.

\subsection{Nutrient Diffusion and Digestive System}

\begin{figure}[ht!]
\centering
\includegraphics[width=6.0in]{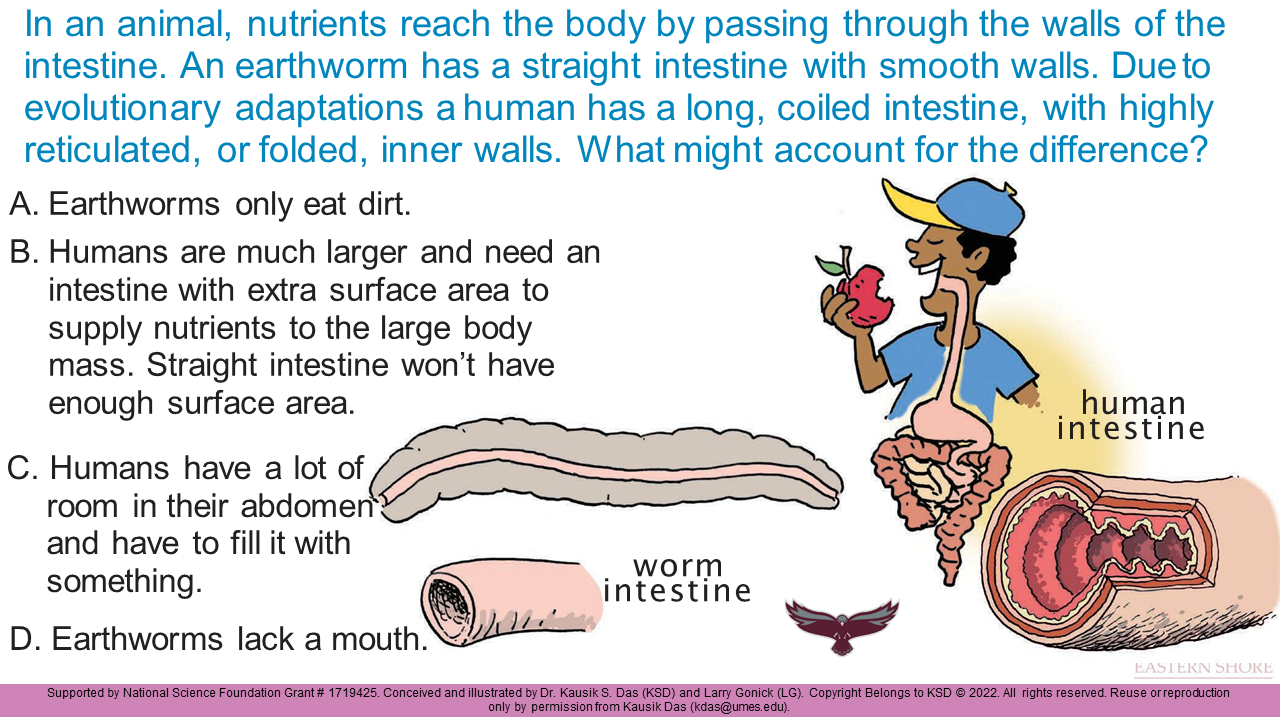}
\caption{\textbf{Surface features and nutrient absorption.} This question explores the differences in intestine structure between humans and earthworms, focusing on different surface area adaptations for efficient nutrient absorption in larger animals.}
\label{Intenstine}
\end{figure}

When considering the evolutionary adaptations related to nutrient absorption in varying species sizes, we observe a similar fascinating trend. Similar to oxygen absorption, nutrients are also absorbed through the walls of the digestive organs. Therefore, requirement of varying nutrient needs often translates to obtaining the right surface area of the nutrient absorbing digestive walls. Our observation is that larger species often develop more complex and efficient digestive systems to handle their increased nutritional needs as they have larger number of cells to feed. This includes longer or more specialized digestive tracts enabling larger surface area for nutrient absorption, enabling them to maximize nutrient extraction and energy efficiency. In Fig.~\ref{Intenstine}, we demonstrated that nutrient absorption in animals is facilitated by the intestinal wall, with different species exhibiting varying structural adaptations. Earthworms, for example, have a straight intestine with smooth walls, sufficient to absorb nutrients for their size. In contrast, much larger animals like humans possess a long, coiled intestine with intricate, folded inner walls. This complex structure provides the necessary increased surface area for absorbing nutrients essential for sustaining a larger body mass. A straight intestine, like that of an earthworm, would not offer enough surface area for the nutritional needs of a human like larger animals. Therefore, mutations that are supportive of this type of increased effective surface area of the digestive tracks would increase the probability of their survival. Such adaptations are crucial for supporting the larger body sizes and higher energy demands of these species, showcasing how evolution tailors physiological processes like digestion to the specific needs of different organisms based on their size.

During a lecture discussion on intestinal adaptations by one of the authors (KD), a thought-provoking inquiry emerged from the audience: Despite their similar body shapes, how does a large snake, such as a python, manage to survive when it is assumed they share a similar intestinal structure with an earthworm? 
Upon investigation we found that while superficially resembling an earthworm, a python's intestines are not smooth but intricately shaped, akin to larger animals~\cite{secor2008digestive}. This finding reinforced our understanding of the relationship between growth, form, and function, as governed by basic scaling laws.

\subsection{Food Habits: Energy Intakes}

The ecological principle that relates an animal's size to its metabolic rate and dietary needs is well-documented. Every animal loses body heat via radiative cooling according to the Stefan-Boltzmann law:
\begin{equation}\label{SB}
P = \epsilon \sigma A T^4,
\end{equation}
where \( \epsilon \) represents the emissivity coefficient of the body surface, a measure of how effectively a surface emits energy as thermal radiation compared to a black body. The emissivity value ranges from 0 to 1, where 1 denotes a perfect black body, an ideal emitter. The rest of the symbols retain their meaning: \( P \) is the power radiated, \( \sigma \) is the Stefan-Boltzmann constant, \( A \) is the surface area, and \( T \) is the temperature in kelvins.

To illustrate that this radiative cooling can be a significant part of the energy budget of an animal, we estimate the energy lost by a human during a day (24 hours) with surface area $A = 2$ m$^2$ of the skin --- which has an emissivity of $\epsilon = 0.98$ \cite{charlton2020effect} --- and internal temperature $T = 305 K$ compared with $300 K$ for the environment. The resulting net loss of energy (received from the environment compared with that emitted by the body) is equivalent to 1250 Calories, a significant part of the daily energy intake. This explains why animals, especially ones that live in cold climates, need adaptations that prevent heat loss, compensate by increasing energy intake, or both

Smaller animals, like mice and shrews, lose body heat relatively rapidly due to their large surface area relative to their body weight. Consequently, they need frequent energy-rich meals high in protein and fat to maintain their body temperature (in addition to other bodily functions). Other small rodents, such as chipmunks and hamsters, also have high surface area to volume ratios, requiring frequent feeding to maintain their energy levels. They consume seeds, nuts, fruits, and small insects --- high-calorie foods --- to obtain the nutrients and calories needed to mitigate their heat loss from their body surfaces.

Small birds, like hummingbirds, and small insects, such as honeybees and butterflies, consume nectar containing high-calorie sugar content almost constantly to support their fast metabolism and high heat loss. Small bats also have high metabolic rates and a high surface area to volume ratio, causing them to lose heat rapidly and necessitating frequent feeding. Depending on the species, bats consume insects, fruits, or nectar, i.e, high-calorie foods, which provide the necessary energy to sustain their high heat loss.

This principle applies to small insects like mosquitoes as well. An interesting fact about mosquitoes is that plant nectar is an essential food source for adult mosquitoes of both sexes~\cite{barredo2020not}. Male mosquitoes especially rely on nectar, needing frequent sugar intake to stay alive. This aspect of their diet is often overlooked because blood-sucking is more well-known due to its role in spreading diseases. However, blood is not a high-calorie diet, and feeding on blood alone is not sufficient to mitigate the high rate of heat loss due to their large surface area relative to their body weight. Some types of mosquitoes feed only on plants, while those that drink blood, like female mosquitoes, also need high-calorie sugar. If male mosquitoes do not get sugar, they usually die within four days after becoming adults. Without the energy from sugar, they cannot fly or mate successfully. Female mosquitoes are relatively less affected by the lack of sugar, but their chances of living longer and laying eggs decrease without enough energy reserves. Therefore, out of all those mutations, the ones that were optimum for their food habits were able to survive. This principle also explains why very small creatures are rarely found in extremely cold regions, such as the Arctic. In cold weather, the rate of heat loss from their bodies is even greater.

In contrast, large animals like elephants have a much lower surface area relative to their volume. Rate of radiative heat loss from their body is relatively low and thus require less energy per unit of body mass. Elephants feed on large quantities of low-calorie plant material, such as leaves, twigs, and bark, which supports their size without the need for high-calorie food.

Marine animals also follow this pattern. Blue whales, despite being the largest animals on Earth, feed on small low calorie density krills. In contrast, Orcas, although part of the whale family, are significantly smaller than blue whales, typically measuring about a quarter to a third of their length. This size difference is also evident in their dietary habits; Orcas consume high-calorie foods such as fish, seals, and occasionally larger whales, a diet that helps compensate for the higher relative heat loss from their body surface due to their smaller size. Other smaller marine predators, like tuna also expend more energy for their size and thus hunt larger, more calorie-rich fish.

These examples illustrate that the size of an animal significantly influences its dietary needs and behavior. Genetic mutations that support these traits become fit to survive, otherwise they go extinct, a concept that is crucial for understanding the dynamics of ecosystems and the evolutionary adaptations of species.

\subsection{Load Bearing and Agility}

The physics of load bearing involves understanding how structures support and distribute various loads through their components, ensuring stability and integrity. At the heart of this is the interplay between forces, material properties, and geometry. For example, in load-bearing walls, beams, and columns, the ability to sustain loads without collapsing is governed by principles such as stress, strain, and equilibrium. A key aspect is the distribution of stress, which is the internal force per unit area within materials. The fundamental equation for stress ($\sigma$) is given by:

\begin{equation}\label{Stress}
\sigma = \frac{F}{A},
\end{equation}
where $F$ is the applied force and $A$ is the cross-sectional area over which the force is distributed. Ensuring that stress does not exceed the material's strength is critical in load-bearing design, allowing structures to safely support the required loads without risking fractures, which is crucial in various applications, from residential buildings to bridges and skyscrapers.

\begin{figure}[ht!]
\centering
\includegraphics[width=6.0in]{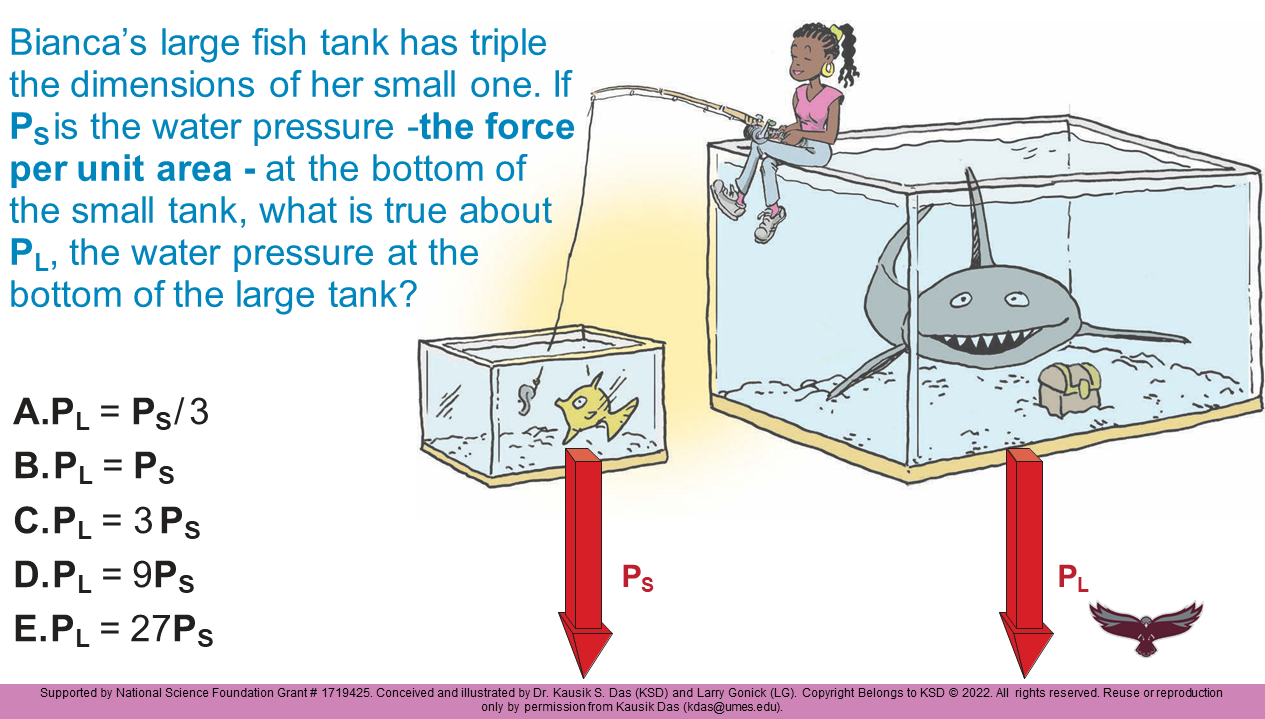}
\caption{\textbf{Scaling of pressure with object size.} Bianca's larger fish tank has dimensions three times those of her small one. This question examines the water pressure at the bottom of the large tank compared to the small tank, illustrating the impact of length scaling on pressure ($\mathbf{P}_L = 3 \mathbf{P}_S$). }
\label{Fish_Tank_Pressure}
\end{figure}

\begin{figure}[ht!]
\centering
\includegraphics[width=6.0in]{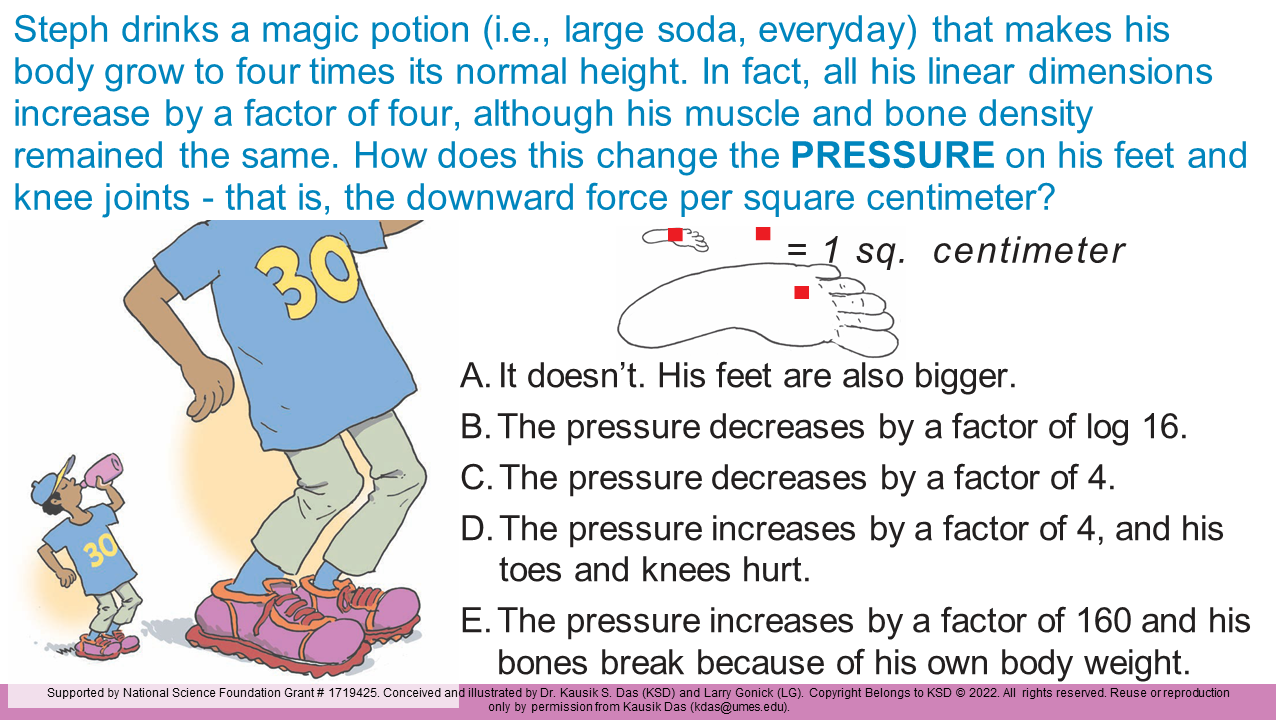}
\caption{\textbf{Scaling of pressure with a human body size.} Steph grows to several times his normal height and width. This question examines how this change affects the pressure on his feet and knee joints. Since body mass (and weight) scales as $L^3$, while cross sectional area scales as $L^2$, pressure increases linearly with body size $L$, just as in Fig.~\ref{Fish_Tank_Pressure}.}
\label{Foot_Pressure}
\end{figure}


\begin{figure}[ht!]
\centering
\includegraphics[width=6.0in]{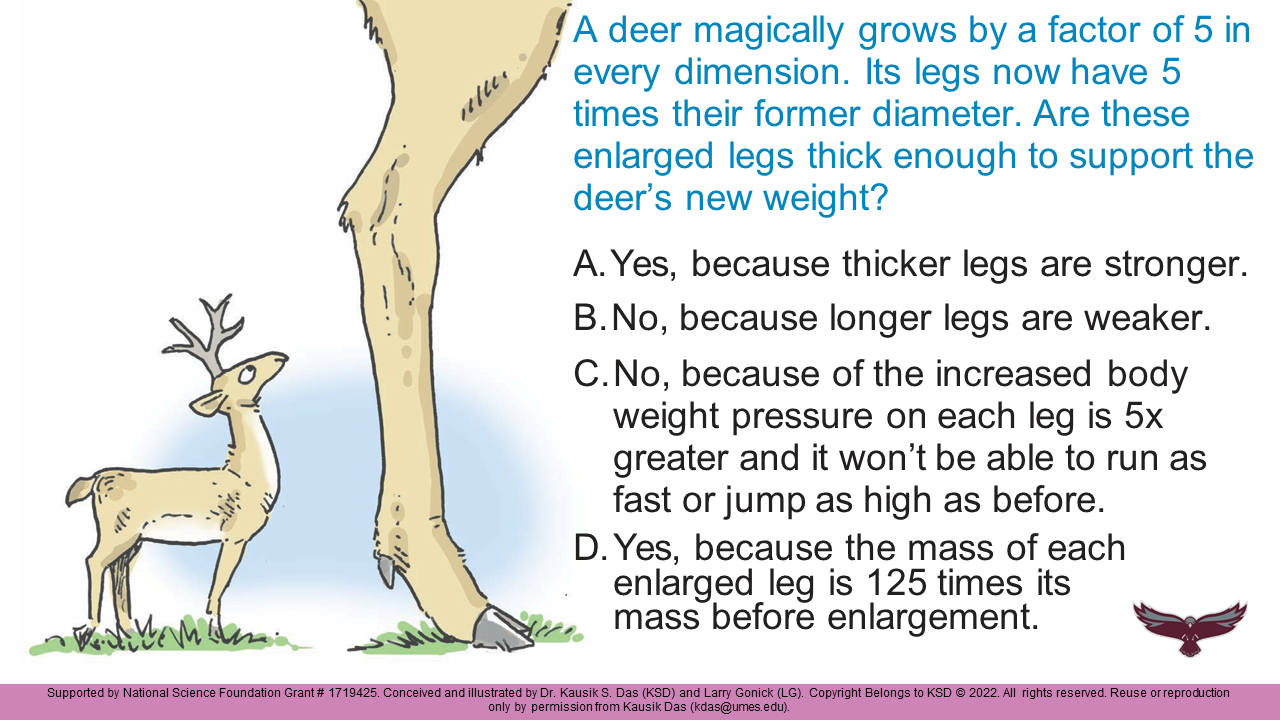}
\caption{\textbf{Scaling of pressure with animal body size and its role in evolution.} Just as in Figs.~\ref{Fish_Tank_Pressure}-\ref{Foot_Pressure}, increasing the size of a deer isotropically in all dimensions will proportionally increase the (cross-sectional) pressure. For example, if a deer scales up five times in every dimension, its weight would become 125 times its original weight, while the cross-sectional area of its leg bones would increase only 25 times. Consequently, the leg bones would experience five times more pressure than before, making it difficult for the massive deer to survive. Such a genetic mutation is therefore unlikely to withstand the test of time.}
\label{Deer}
\end{figure}

Figure~\ref{Fish_Tank_Pressure} illustrates the correlation between the size of a fish tank and the pressure exerted on its base, which is analogous to how weight gain affects lower body joints, as seen in Fig.~\ref{Foot_Pressure}. These visual examples serve a dual purpose: a) they elucidate the concept that a higher BMI (Body Mass Index) puts more pressure on the bones supporting the body's weight, and b) they promote the importance of maintaining a healthy diet to prevent related health issues, wherever possible.

Within the domain of evolutionary biology, the scaling up of an organism's size has significant biomechanical implications. This concept is exemplified in Fig.~\ref{Deer}, which indicates that an increase in the body size of a deer, preserving its shape, results in its volume or weight growing at a cubic rate relative to its characteristic length, while the bone's cross-sectional area increases only at a square rate. This means the stress or pressure on the femur bone increases (scales as $\sim L$), potentially reducing the deer's agility, an important characteristic crucial for escaping predators. To adapt, such species might evolve (via genetic mutations) thicker legs with larger cross-sectional areas, altering their original form, as observed in large mammals like elephants and rhinos, or increase bone density, as seen in large flightless birds or ratites such as ostriches, emus, rheas or kiwis. These adaptations help to distribute the increased weight more effectively, preserving mobility and structural integrity, thereby making them better suited to survive.

\subsection{Interaction with the Environment: Water}

The influence of the surface area-to-volume ratio ($\rho$) extends to the physics of how animals interact with their environment, including water. For instance, insects have a high surface area relative to their volume, which means that when they are drenched in water, surface tension can trap them. Due to surface tension's adhesive properties water cling to the insect's body and spread, forming a thin layer of liquid over their body surface. This added weight from the water layer relative to the animal's body weight determines whether the animal will get trapped in water or not.
Since A/V$\rightarrow\infty$, as $L\rightarrow 0$, as discussed in Fig.~\ref{A_V_Asymptotic}, the weight of the added water on the body surfaces of small insects becomes much heavier in comparison to their own body weight. This makes it almost impossible for small insects like fruit flies to escape the liquid trap when they get wet. Slightly larger but still small mammals, like mice, experience a similar struggle. Fur on the surface of their bodies can absorb water, adding more weight that is substantial compared to their small size, thus impeding their ability to escape from water easily. Larger animals, such as humans, have a much lower value of $\rho$ due to their large characteristic length scale. When we take a dip in a lake or swimming pool and come out, the water layer that clings to our body surface adds a negligible amount of weight in comparison to our own body weight, making it easier for us to carry this water and move without difficulty. The same principle applies to even larger animals like hippos and polar bears, where the weight of water on their massive bodies is minuscule relative to their total body weight, allowing them to move in and out of water with relative ease.

These examples demonstrate that the A/V ratio is a crucial factor in the biomechanics of different species and their ability to interact with their environments, especially regarding their encounters with water. These constraints can also explain why most insects evolve to drink while sitting on the edge of a water source. Mutations and adaptations leading to this habit give them a chance of survival by helping them avoid getting wet and trapped by water.

\section{Transition from Natural to Technological Evolution}
\label{sec:technology}

\subsection{Aviation}

Birds of the same feather flock together, whereas birds of different species do not, as they have varying body weights, wingspans, and different cruising speeds. It is therefore interesting to investigate how birds’ surface area to volume ratio affects their cruising speeds and their evolutionary traits. A bird's wing loading is defined as the ratio of its weight to the surface area of its wings, which is approximately proportional to the volume-to-surface area ($V/A = 1/\rho$) ratio.

To achieve flight, a flow past body like an airfoil or a birds wings must generate lift or upward force that is at least equal to its own weight \(W\). From the physics of aviation we know that lift force \(F\) is given by the equation:
\begin{equation}\label{Lift}
 F = W = C v^2 A,
\end{equation}
where: \(C\) is a function of the lift coefficient, which is related to the angle of attack and air density, \(v\) is the cruising speed, and \(A\) is the wing area.

\begin{figure}[ht!]
\centering
\includegraphics[width=6.0in]{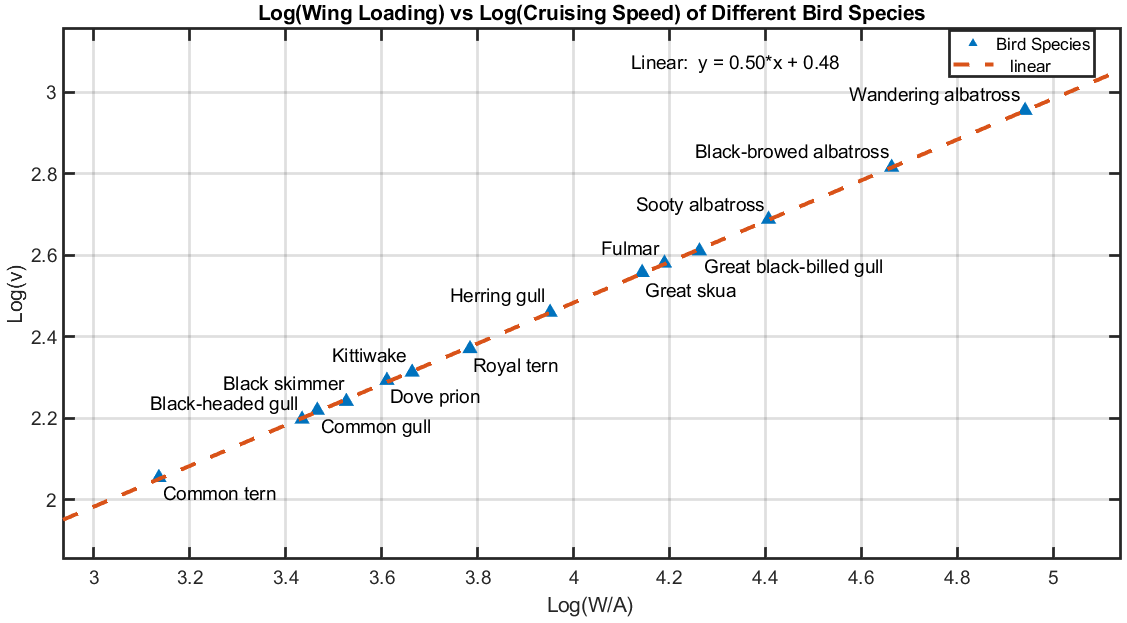}
\caption{\textbf{The relationship between log-transformed wing loading $\log(W/A)$ and log-transformed cruising speed $\log(v)$ for different bird species.} Each point represents a specific bird species, and the linear regression line (dashed orange line) shows a positive correlation between wing loading and cruising speed, with the equation \( y = 0.50x + 0.48 \). The data spans a range of bird species from the common tern, with low wing loading and cruising speed, to the wandering albatross, with high wing loading and cruising speed. This demonstrates the universality of the scaling relationship between these physical characteristics, reflecting a common evolutionary adaptation among diverse bird species to optimize their flight performance and survive.}
\label{Birds_log}
\end{figure}

Since birds' weight depend on their sizes, we can see that the wing loading of different birds can be written as:
\begin{equation}\label{WingLoading}
\frac{W}{A} \sim \frac{V}{A} \sim v^2,
\end{equation}
where, \(W\) is the weight of the individual birds. This means that the ratio of a bird's volume to its wing area (\(V/A\)) is proportional to the square of its cruising speed \(v^2\). Thus, for example, according to this analysis, if one bird's wing loading is four times that of another bird, then the cruising speed of the first bird should be twice as fast as the second! Furthermore, if we take the logarithm of both sides of the equation Eq.~(\ref{WingLoading}), we get:
\begin{equation}\label{loglog}
  \log(v) \sim \frac{1}{2} \log(W/A),
\end{equation}
Equation~\eqref{loglog} also reveals that, from the physics of lift perspective, the slope of the linear \(\log(v)\) vs. \(\log(W/A)\) graph should be close to 0.5. Interestingly real flight data of different bird species, which we collected from Ref.~\cite{tennekes2009simple}, conform to this trend (see table \ref{table:wing_loading_cruising_speed} and Fig.~\ref{Birds_log}).

\begin{table}[ht]
\centering
\begin{tabular}{|p{1.8in}|p{0.8in}|p{1.0in}|p{1.0in}|p{1.0in}|}
\hline
\textbf{Bird Species}& \textbf{Weight (kg)} & \textbf{Wing Loading (N/m$^2$)} & \textbf{Cruising Speed (m/s)} & \textbf{Cruising Speed (mph)} \\ \hline
Common tern                    & 0.11                 & 23                              & 7.8                           & 18                             \\ \hline
Dove prion                     & 0.20                 & 37                              & 9.9                           & 22                             \\ \hline
Black-headed gull              & 0.28                 & 31                              & 9.0                           & 20                             \\ \hline
Black skimmer                  & 0.35                 & 34                              & 9.4                           & 21                             \\ \hline
Common gull                    & 0.40                 & 32                              & 9.2                           & 21                             \\ \hline
Kittiwake                      & 0.41                 & 39                              & 10.1                          & 23                             \\ \hline
Royal tern                     & 0.45                 & 44                              & 10.7                          & 24                             \\ \hline
Fulmar                         & 0.80                 & 66                              & 13.2                          & 30                             \\ \hline
Herring gull                   & 0.85                 & 52                              & 11.7                          & 26                             \\ \hline
Great skua                     & 1.41                 & 63                              & 12.9                          & 29                             \\ \hline
Great black-billed gull        & 1.75                 & 71                              & 13.6                          & 31                             \\ \hline
Sooty albatross                & 2.50                 & 82                              & 14.7                          & 33                             \\ \hline
Black-browed albatross         & 3.40                 & 106                             & 16.7                          & 38                             \\ \hline
Wandering albatross            & 5.00                 & 140                             & 19.2                          & 43                             \\ \hline
\end{tabular}
\caption{Wing loading vs cruising speed of different bird species \cite{tennekes2009simple}.}
\label{table:wing_loading_cruising_speed}
\end{table}

When comparing the cruise speeds of different birds, such as the Dove Prion and the Kittiwake, as well as the Fulmar and the Great Skua, it becomes evident that wing loading plays a crucial role in determining their natural cruising speed, rather than weight alone. The Dove Prion, weighing 0.20 kg, has a wing loading of 37 N/m² and a cruising speed of 9.9 m/s. Despite the Kittiwake being significantly heavier (nearly double) at 0.41 kg, it has a similar wing loading of 39 N/m², resulting in a very similar cruising speed of 10.1 m/s. Similarly, the Fulmar, with a weight of 0.80 kg and a wing loading of 66 N/m², cruises at 13.2 m/s. In contrast, the Great Skua, which is much heavier at 1.41 kg, has a similar wing loading of 63 N/m² but a close cruising speed of 12.9 m/s. These examples illustrate that even with considerable differences in weight, wing span and other features, birds can have very similar cruising speeds if their wing loadings are close, highlighting the importance of volume-to-surface-area ratio in determining flight performance. 

This aspect of cruising speed as a function of wing loading can also be observed in the technological evolution of aircrafts. From Eq.~(\ref{Lift}) we know that to maintain lift for a cruising heavier airborne object we need to increase the surface area of the wing proportionally. However, operational constraints prevents us from increasing the wing surface area indiscriminately. For instance, the Boeing 777, with a wing area only six times larger than that of the smaller Fokker F27, can carry a weight more than 12 times greater! To understand this, we need to consider the principles of aerodynamics as discussed earlier in Eq.~\eqref{Lift}. For the Boeing 777, achieving sufficient lift with a relatively modest increase in wing area involves flying at a significantly higher cruising speed compared to the Fokker F27. This is because the lift generated by the wings is proportional to the square of the velocity. In simpler terms, even though the Boeing 777's wing area is only six times larger, it compensates for its much greater weight by flying faster. This increased speed enhances the lift generated by the wings, allowing the larger aircraft to stay airborne.

Therefore, the evolution of aircraft design also demonstrates how engineers optimize various factors such as wing area and cruising speed to achieve the necessary lift for heavier aircraft with larger volumes. This balance allows larger aircraft like the Boeing 777 to carry significantly more weight without requiring disproportionately larger wings, as seen in the evolutionary adaptations of the birds (Fig.~\ref{Birds_log}).

\subsection{Charge Storage}
Efficient electrical energy storage is a grand challenge in modern technological progress. Efficient energy storage solutions, such as batteries and supercapacitors, depend on optimizing the charge storage density or specific capacity. This principle is vital for enhancing the performance and sustainability of technologies like electric vehicles (EVs), flying taxis, drones, and renewable energy systems. Advanced high energy density batteries and supercapacitors can store more energy while keeping its weight low, improving the range and efficiency of EVs, flying taxis, robots etc. Similarly, drones, battery powered underwater vehicles etc. also benefit from longer life times due to improved energy density. In renewable energy systems, optimized charge storage allows for better integration of solar and wind power into the grid without making it bulky and heavy, stabilizing energy supply and reducing fossil fuel dependence. Therefore, understanding the evolution of energy storage technologies is important for advancing modern technology and combating climate change.

The capacitance \( C \) of a double plate capacitor is given by the equation:

\[ C = \frac{\varepsilon A}{d} \]

where:
\( \varepsilon \) is the permittivity of the dielectric material between the plates,
\( A \) is the surface area of one of the plates,
\( d \) is the distance between the plates. From this equation, we can see that the capacitance \( C \) is directly proportional to the surface area \( A \) of the plates. This relationship indicates that increasing the surface area of the plates will increase the capacitance, allowing the capacitor to store more charge.
\begin{figure}[ht!]
 \centering
\includegraphics[width=6.0in]{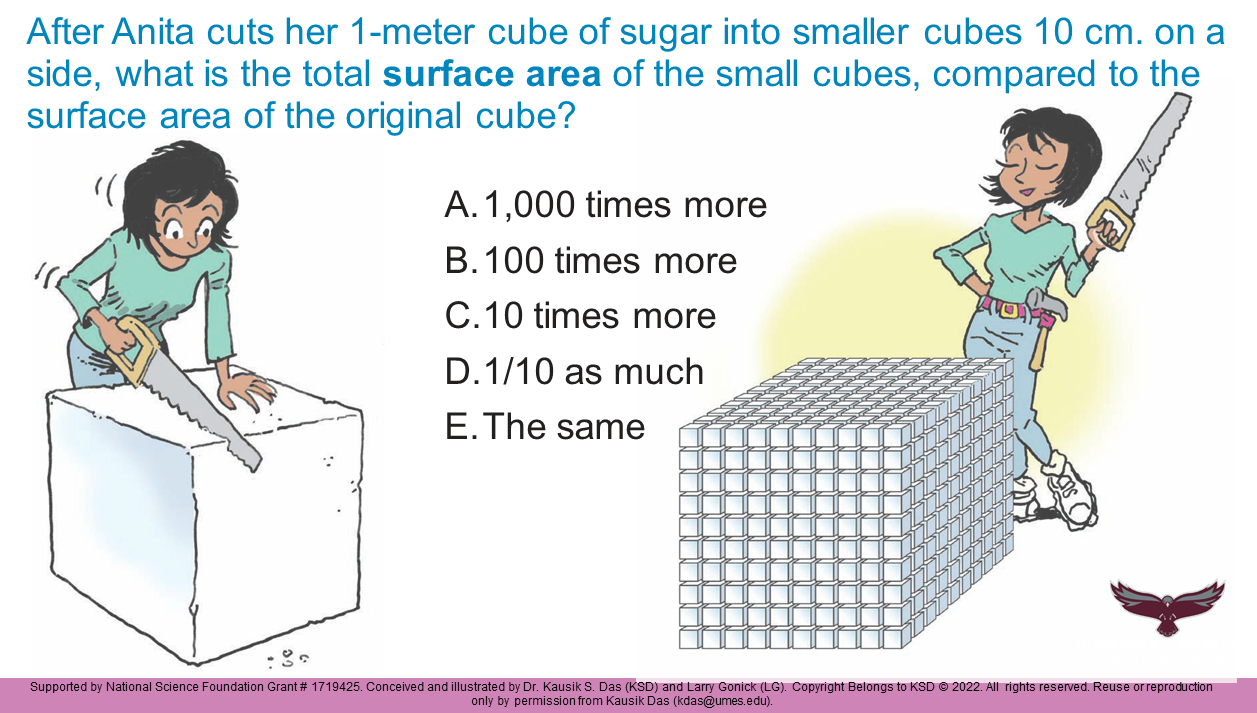}
\caption{\textbf{Surface area and fragmentation}. The question asks for the total surface area of the smaller cubes compared to the surface area of the original cube. The visual depicts Anita cutting the large cube into smaller cubes, prompting the calculation of the resulting increase in total surface area. This exercise highlights the dramatic increase in surface area when an object is fragmented into smaller pieces.}
\label{Lucky_Anita}
\end{figure}
In the context of supercapacitors, this principle is taken to an advanced level. Supercapacitors, also known as electrochemical capacitors or ultracapacitors, utilize materials with extremely high surface areas to maximize capacitance. Nano materials and 2D materials like graphene are particularly effective because of their large surface-to-volume ratios~\cite{el2012laser}.

\begin{figure}[ht!]
\centering
\includegraphics[width=6.0in]{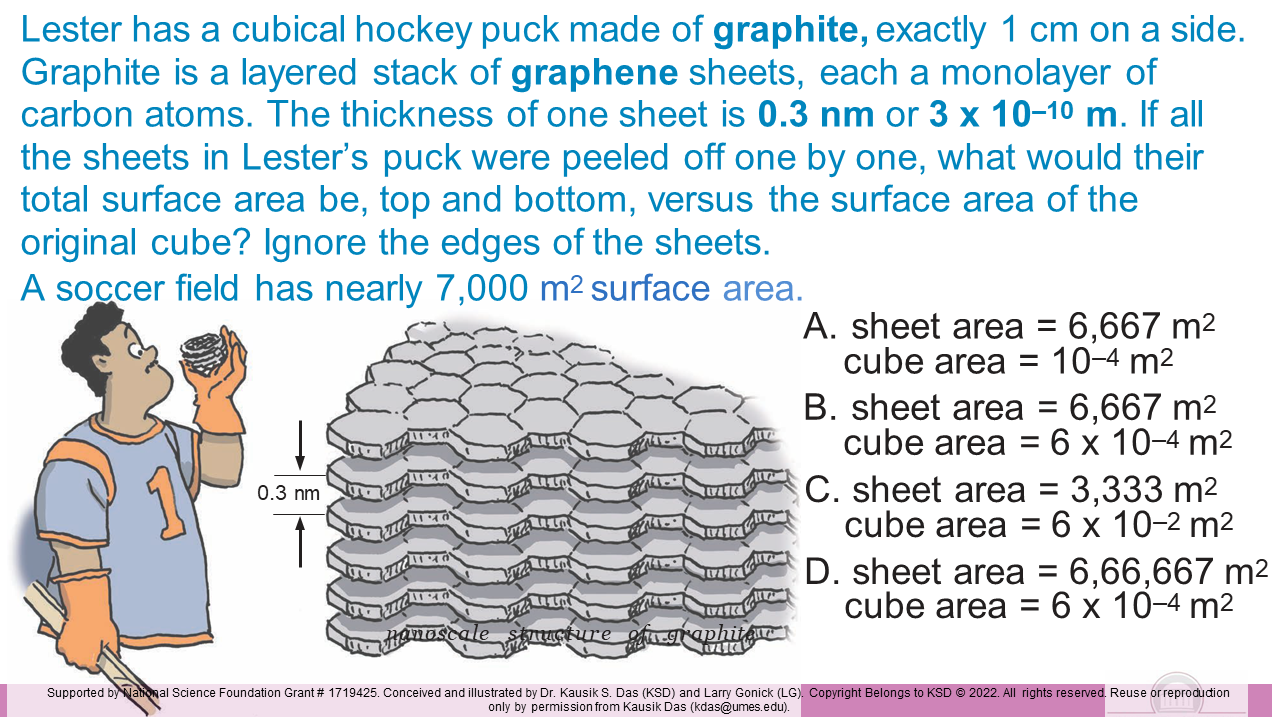}
\caption{\textbf{The surface area of Graphite vs the total surface area of the graphene sheets that make it up.} The question asks for the total surface area resulting from the graphene sheets that are peeled off one by one, compared to the surface area of the original graphite cube (ignoring the edges of the sheets). This total surface area is compared to the surface area of a soccer field, nearly 7,000 $m^2$. }
\label{Graphene_Stack}
\end{figure}

\begin{figure}[ht!]
\centering
\includegraphics[width=6.0in]{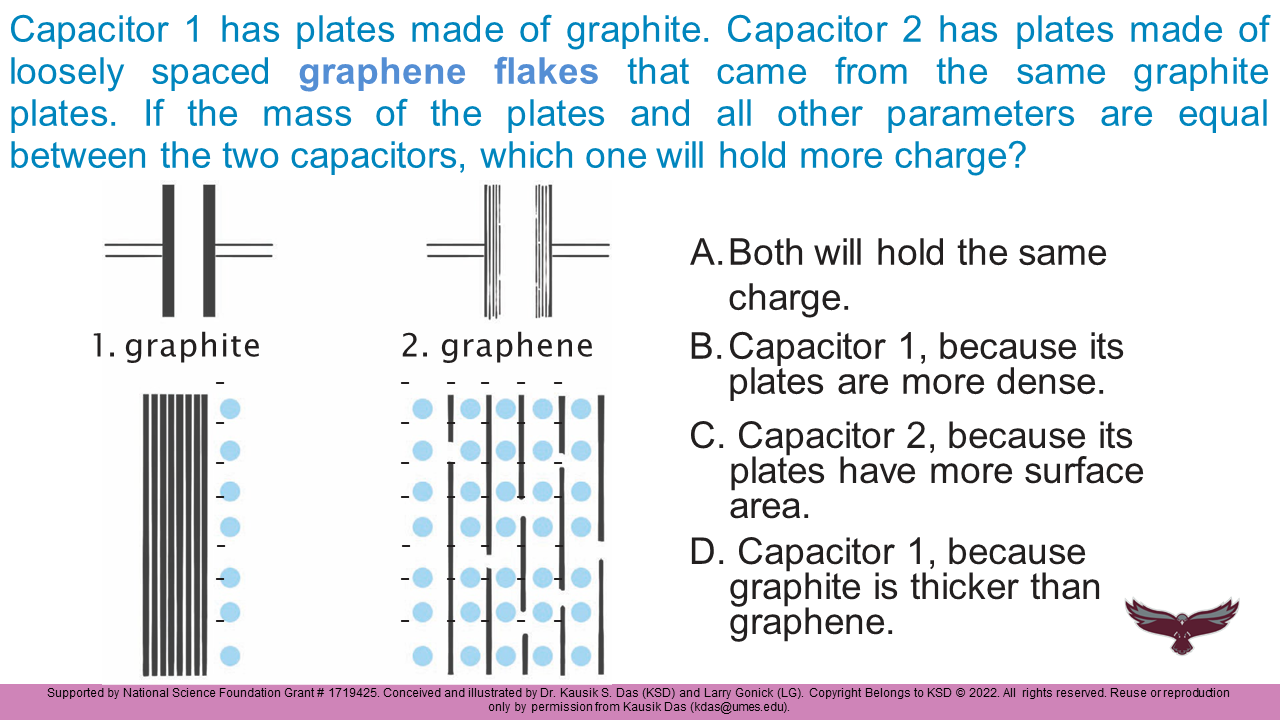}
\caption{\textbf{Supercapacitors and surface area.} Building on Figs.~\ref{Lucky_Anita}-\ref{Graphene_Stack}, this figure illustrates two capacitors with different plate materials: Capacitor 1 has electrodes made of graphite, while Capacitor 2 has electrodes made of loosely spaced graphene flakes derived from the same graphite plates. This shows the impact of surface area on the capacitance, or the charge-holding capacity, of capacitors.}
\label{Graphene_Super_Cap}
\end{figure}
Graphene, for example, is a single layer of carbon atoms arranged in a two-dimensional honeycomb lattice. This structure provides a vast surface area in a very small volume, significantly enhancing the ability to store charge in small space. When these materials are used in supercapacitors, the electrodes can have much larger effective surface areas compared to traditional materials, leading to a dramatic increase in capacitance.

Therefore, by using nano materials and 2D materials, supercapacitors exploit the direct proportionality between surface-to-volume ratio and capacitance to achieve higher energy storage capacities. This enhancement allows supercapacitors to store a lot of charge in a small volume and deliver it more efficiently, with faster charge and discharge cycles, making them ideal for applications requiring rapid energy bursts and high power densities.

\subsection{Heat Transfer}
Achieving efficient heat transfer is crucial in many technological applications. For instance, while we are now capable of manufacturing high-performing computer chips, the lack of efficient heat transfer mechanisms hinders the growth of the industry. The applications of heat transfer \cite{das2010dynamics,das2000onset} extend to various fields, including aerospace science, microelectromechanical systems (MEMS), and compact heat exchangers. Massive data centers are even being submerged underwater in a desperate attempt to cool them by transferring heat to the surrounding sea. For a composite system involving conduction through a material, convection from the surface to a fluid, and radiation to the surroundings, the combined heat transfer rate might look like this:

\begin{equation}
Q_{\text{total}} = kA \frac{dT}{dx} + hA(T_s - T_\infty) + \sigma \epsilon A (T_s^4 - T_{\text{surroundings}}^4)
\end{equation}

where:

\begin{itemize}
  \item $k$ is the thermal conductivity of the material,
  \item $\frac{dT}{dx}$ is the temperature gradient within the material,
  \item $h$ is the convective heat transfer coefficient,
  \item $T_\infty$ is the temperature of the fluid,
  \item $\sigma$ is the Stefan-Boltzmann constant ($5.67 \times 10^{-8}$ W/m²K$^4$),
  \item $\epsilon$ is the emissivity of the surface,
  \item $T_s$ is the surface temperature,
  \item $T_{\text{surroundings}}$ is the temperature of the surrounding environment.
\end{itemize}

It is clear that every form of heat transfer rate is proportional to the surface area $A$ of the object. Therefore, materials with a higher surface area to volume ratio will be more efficient in transferring heat compared to those with a lower ratio. This principle explains why micro/nanomaterials are often added to heat transfer fluids or why surfaces with specific topography, such as heat transfer fins, are engineered to significantly enhance heat transfer. By using micro/nanoparticles, we significantly increase the surface area to volume ratio, enhancing heat conduction. For example, in nanofluids --- fluids containing nanoparticles --- the large surface area of nanoparticles facilitates significantly higher heat transfer compared to conventional fluids~\cite{xie2010intriguingly,sackmann2014present}. This is because the higher surface area allows more interaction between particles and the surrounding fluid, leading to improved thermal conductivity even using a small volume of material, making it and efficient mechanism. These advancements are essential not only for cooling high-performance electronics but also for applications in renewable energy systems, advanced manufacturing, and thermal management in various industries.

Therefore, optimizing the surface area to volume ratio is essential for advancing modern technology and combating challenges related to heat dissipation. By leveraging micro/nanomaterials and engineered surface topographies, we can control heat transfer significantly, supporting the development of more efficient and powerful technological solutions.

\begin{figure}[ht!]
\centering
\includegraphics[width=6.0in]{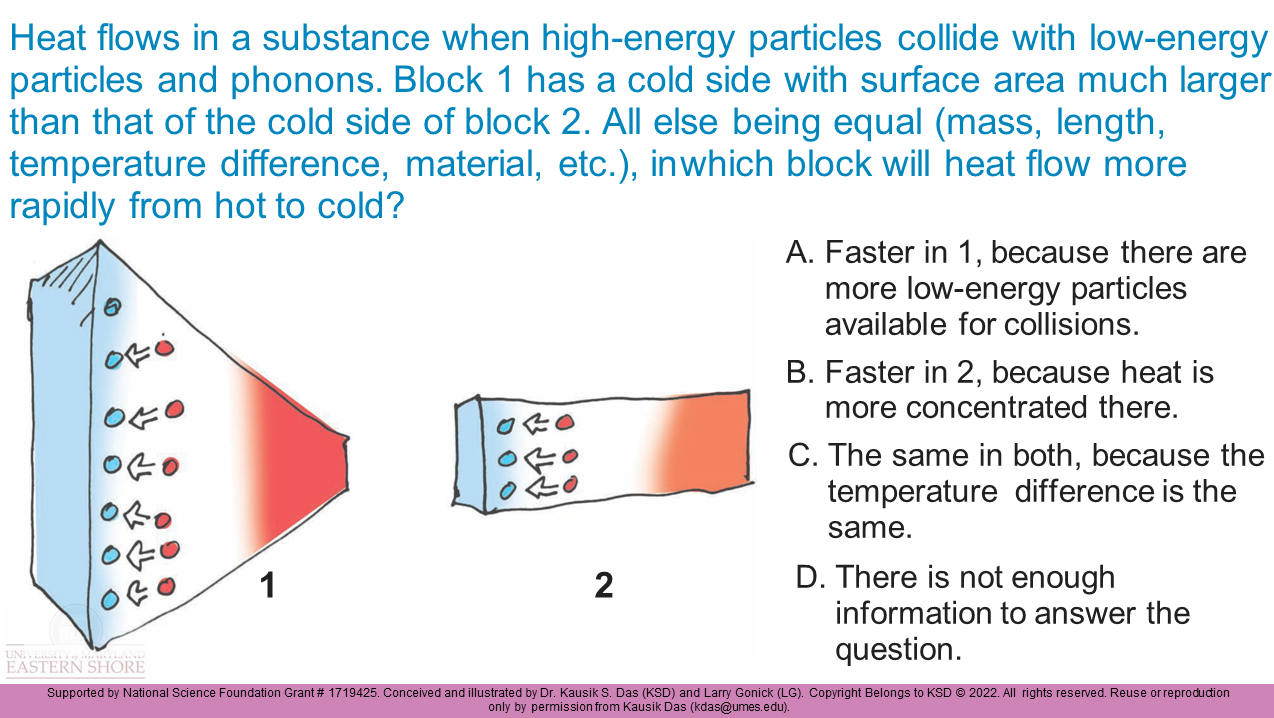}
\caption{\textbf{Heat conduction and cross-sectional area.}  This question and associated visual illustrates how surface area impacts the rate of heat transfer. More cross-sectional area leads to more collisions between high and low energy particles (and phonons).}
\label{Heat_Conduction1}
\end{figure}

\begin{figure}[ht!]
\centering
\includegraphics[width=6.0in]{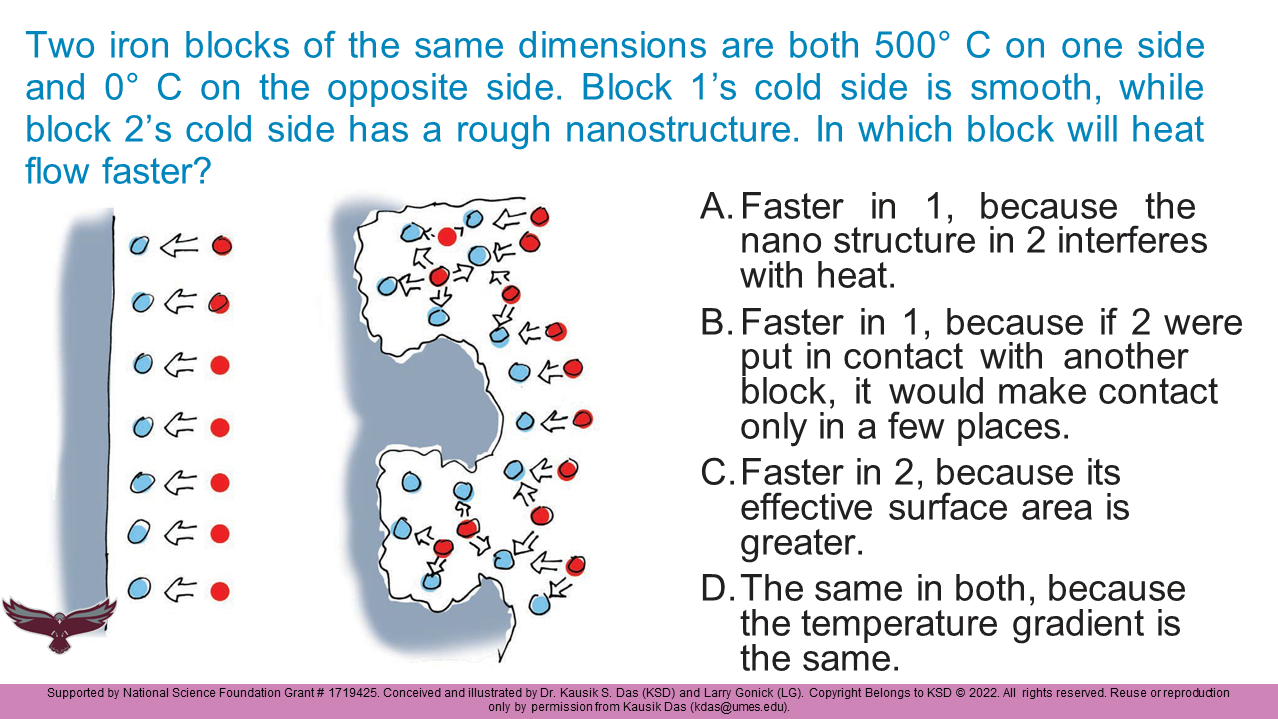}
\caption{\textbf{Heat transfer and surface features.} Since rough surfaces have higher surface areas --- as demonstrated in Figs.~\ref{Chess}-\ref{Intenstine} --- this figure illustrates the enhancement in the rate of heat flow across a piece of metal with a rough surface, as compared with a smooth one. 
}
\label{Heat_Conduction2}
\end{figure}

\subsection{Microfluidics}
Microfluidics has the potential to impact many areas of life, including science, medicine, and industry~\cite{battat2022outlook,sackmann2014present}. In the context of microfluidics, the impact of surface area to volume ratio is profound, especially in low Reynolds number regimes where mixing is dominated by laminar flow. The article  illustrates how manipulating the surface area to volume ratio through innovative boundary conditions can significantly enhance microfluidic mixing.

In microchannels, the ratio of the surface area to the volume becomes very large due to the small characteristic length scales. This high ratio means that surface effects, such as boundary conditions, have a substantial impact on fluid dynamics. By introducing two-dimensional hydrophobic slip patterns on the floor of microchannels~\cite{ouro2022boundary,nazari2020surface,foroughi2012immiscible,barnes2021plasma}, Das and his colleagues were able to induce chaotic advection and enhance mixing efficiency. These hydrophobic patterns create a discontinuous change in wall boundary conditions, forcing the fluid streamlines to adjust instantaneously, leading to stretching, folding, and recirculation of the fluid. This method improves mixing without the need for complex and costly three-dimensional micro-patterning techniques. The enhanced mixing achieved through optimized surface area to volume ratios in microfluidic systems has broad implications. It can improve the performance of lab-on-a-chip devices, chemical reactors, and biological assays by enhancing reaction rates, improving detection sensitivity, and reducing sample and reagent consumption. This approach represents a significant step forward in microfluidic technology, offering practical and cost-effective solutions to current challenges.

\section{Conclusion}
In conclusion, the interplay between evolutionary biology and the principles of physics, particularly the surface area to volume (A/V) ratio, reveals profound insights into the adaptations of various species. As organisms grow in size, their physiological structures and functions must adapt to maintain efficiency and survival. This study highlighted the critical role of the A/V ratio in influencing respiratory and digestive systems, heat regulation, load-bearing capacities, and interactions with the environment.

Smaller organisms, due to their higher A/V ratios, face challenges such as rapid heat loss and difficulties in nutrient and oxygen absorption. These challenges necessitate frequent feeding and specialized respiratory structures to maintain metabolic functions. Conversely, larger animals have evolved complex structures like lungs and extensive blood vessel networks to enhance oxygen uptake, allowing them to meet their heightened metabolic demands efficiently.

The principle of form following function, underscored by the A/V ratio, extends to technological advancements as well. In fields like aviation, energy storage, and heat transfer, optimizing surface area relative to volume enhances performance and efficiency. For instance, the design of aircraft wings and the development of supercapacitors leverage these principles to achieve superior functionality.

Integrating these concepts into physics education can enrich students' understanding of the interconnectedness of physical and biological sciences. By using real-world examples and engaging teaching methods, such as concept cartoon clicker questions, educators can deepen students' comprehension of abstract physical laws and their applications in biological contexts.

Overall, the optimization of the surface area to volume ratio, driven by genetic mutations and selection by natural physical laws, not only dictates the survival and functionality of living organisms but also propels technological innovation. This comprehensive approach bridges the gap between biology and physics, providing a holistic view of life's adaptations and the advancement of modern technology.

\section{Acknowledgement}
KD would like to acknowledge Kavli Institute of Theoretical Physics (KITP) for their support through the KITP Fellowship program. This support facilitated KD's visit and stay at KITP, enabling the opportunity for many thought provoking discussions. This work was supported in part by the National Science Foundation under Grant No. NSF PHY-1748958, the NSF HBCU-UP Award \# 1719425, and the Department of Education (MSEIP Award \# P120A70068) with MSEIP CCEM Supplemental grant.
\clearpage
\section{References}
\bibliography{BibtexdatabaseKSD}

\end{document}